\documentclass[12pt,preprint]{aastex}

\usepackage{graphics,graphicx}
\usepackage{subfigure}
\usepackage{color}
\usepackage{lscape}

\begin{document}

\renewcommand{\thefootnote}{\fnsymbol{footnote}}

\title{Disentangling AGN and Star Formation in Soft X-rays}

\author{Stephanie M. LaMassa$^{1}$\footnote{S. M. LaMassa is now at Yale University.}, T. M. Heckman$^{1}$, A. Ptak$^{2}$}

\affil{$^1$The Johns Hopkins University
$^2$ NASA/GSFC
}

\begin{abstract}
We have explored the interplay of star formation and AGN activity in soft X-rays (0.5-2 keV) in two samples of Seyfert 2 galaxies (Sy2s). Using a combination of low resolution CCD spectra from {\it Chandra} and {\it XMM-Newton}, we modeled the soft emission of 34 Sy2s  using power law and thermal models. For the 11 sources with high signal-to-noise {\it Chandra} imaging of the diffuse host galaxy emission, we estimate the luminosity due to star formation by removing the AGN, fitting the residual emission. The AGN and star formation contributions to the soft X-ray luminosity (i.e. L$_{x,AGN}$ and L$_{x,SF}$) for the remaining 24 Sy2s were estimated from the power law and thermal luminosities derived from spectral fitting. These luminosities were scaled based on a template derived from XSINGS analysis of normal star forming galaxies. To account for errors in the luminosities derived from spectral fitting and the spread in the scaling factor, we estimated L$_{x,AGN}$ and L$_{x,SF}$ from Monte Carlo simulations. These simulated luminosities agree with L$_{x,AGN}$ and L$_{x,SF}$ derived from {\it Chandra} imaging analysis within a 3$\sigma$ confidence level. Using the infrared [NeII]12.8$\mu$m and [OIV]26$\mu$m lines as a proxy of star formation and AGN activity, respectively, we independently disentangle the contributions of these two processes to the total soft X-ray emission. This decomposition generally agrees with L$_{x,SF}$ and L$_{x,AGN}$ at the 3$\sigma$ level. In the absence of resolvable nuclear emission, our decomposition method provides a reasonable estimate of emission due to star formation in galaxies hosting type 2 AGN.
\end{abstract}

\section{Introduction}

Supermassive black holes (SMBHs) and their parent galaxies co-evolve \citep{kr, magorrian, FM, gebhardt, tremaine, hr}. In particular, observational and theoretical work has established a link between accreting SMBHs (active galactic nuclei, or AGN) and host galaxy star formation. Common mechanisms have been proposed for triggering star formation while fueling SMBH accretion, including galaxy mergers \citep[e.g.][]{Sanders, Hopkins} and secular processes involving gravitational instabilities induced by spiral arms or galactic sized bars \citep[e.g.][]{KK, Cisternas, Schawinski}.

Disentangling star formation from AGN activity becomes a necessary endeavor to investigate SMBH and host galaxy co-evolution. This separation has been explored extensively and has included identifiying optical and ultraviolet (UV) spectral signatures of starburst activity in AGN \citep{Cid}, performing principal component analysis on infrared spectra of AGN \citep[e.g.][]{Buchanan}, and analyzing optical and infrared (IR) diagnostics that parameterize the relative contribution of AGN to star formation, such as ratios of infrared and optical emission lines that indicate ionization field hardness \citep{Genzel, Kewley01, Kauffmann, Treyer, me2, me4} and strength of the polycyclic aromatic hydrocarbon features \citep{Genzel, O'Dowd, me2, me4}. Here we extend the study of the interplay between AGN activity and star formation to the X-ray regime.

Quiescent galaxies emit thermal and non-thermal X-ray emission. Hot gas energized by stellar winds and supernova explosions radiates in soft X-rays (0.5 - 2 keV) whereas X-ray binares and supernova remnants dominate the non-thermal emission above 2 keV \citep[see][for a review]{Fabbiano}. Due to the relatively short life time of high mass X-ray binaries (HMXBs, $\tau < 10^{7}$ yr) and the short delay between starburst and generation of soft X-rays from hot gas, hard and soft X-ray emission can trace the instantaneous star formation rate (SFR). Indeed, X-ray emission from quiescent galaxies is well correlated with radio, infrared, optical and ultraviolet SFR indicators \citep[e.g.][]{Ranalli, Rosa, PS}. Various X-ray SFRs have thus been calibrated in the literature based on the correlation of X-ray emission with far-infrared and radio luminosities \citep{Ranalli}, X-ray emission from HMXBs and cumulative galactic X-ray point sources with far-infrared luminosities \citep{Persic04, Persic07} and X-ray luminosity with infrared plus ultraviolet emission \citep{PS}. However, low mass X-ray binaries, which trace stellar mass (M$_{\star}$) rather than star formation, also contributes to the X-ray emission and becomes the primary source of X-ray radiation in galaxies with low SFR/M$_{\star}$ \citep{Colbert, Lehmer}.

An active nucleus complicates using X-ray emission as a SFR indicator: accretion onto the SMBH dominates emission above 2 keV. Isolating the hard X-ray emission due to star formation then becomes prohibitively difficult when the nucleus is not resolved. However, soft X-ray emission can have comparable contributions from scattered/reflected AGN emission and thermal emission from gas associated with starburst activity when the AGN is obscured or has very low luminosity. Determining the relative contribution of AGN and star formation to the soft X-ray emission can be achieved through high resolution spectroscopy which resolves narrow emission lines. Various diagnostics can be used to distinguish between collisionally and photoionized plasma, such as the width of the radiative recombination continua \citep{LP96} and ratios of forbidden, intercombination and resonance lines in the OVII triplet \citep{PD00}. However, with current X-ray missions (i.e. {\it Chandra} and {\it XMM-Newton}) high resolution spectra are only obtainable through grating observations, necessitating long exposure times on X-ray bright sources to achieve adequate signal to noise. How well can less time-intensive low resolution CCD spectroscopy achieve the goal of disentangling starburst from AGN activity in soft X-rays?

Local obscured AGN, Seyfert 2 galaxies (Sy2s), provide an ideal laboratory to answer this question. As these sources are relatively nearby and the accretion disk is hidden behind an obscuring ``torus,'' circumnuclear starbursts are visible. Hence, both AGN and host galaxy star formation can be modeled. Levenson et al. (2004, 2005) analyzed {\it Chandra} observations of two Sy2/starburst composites (NGC 5135 and NGC 7130) and demonstrated that a thermal component due to stellar processes was necessary to adequately model the soft X-ray emission. We expand this methodology to Sy2s in general to investigate the efficacy of modeling soft X-ray emission with a thermal component to describe host galaxy star formation.

Our analysis focuses on two samples of Sy2s: an optically selected [OIII]5007\AA\ sample \citep{me1} and a mid-infrared selected 12$\mu$m sample \citep{12m}. Using {\it Chandra} and {\it XMM-Newton} observations, we decomposed the soft X-ray emission into a star-forming and an AGN component. For Sy2s observed with {\it Chandra} that have significant host galaxy emission, we are able to resolve the point source associated with the AGN and can therefore remove it and analyze emission soley from the host galaxy, which we ascribe to star formation (L$_{x,SF}$). For the Sy2s observed with {\it XMM-Newton} and the remaining {\it Chandra} sources, we fit the X-ray spectra with a combination of a thermal and power-law model. To account for the presence of X-ray binaries, we scale the soft thermal and power law fluxes using star-forming galaxies from the XSINGS sample as a template (Ptak et al. 2012), obtaining soft X-ray luminosity values associated with star formation (L$_{x,SF}$) and AGN activity (L$_{x,AGN}$). We compare these estimates of L$_{x,SF}$ and L$_{x,AGN}$ with infrared (IR) spectral signatures that accurately describe star formation ([NeII]12.8$\mu$m luminosity, L$_{[NeII]}$,  \citet{Ho, me4} and AGN activity ([OIV]26$\mu$m luminosity, L$_{[OIV]}$, \citet{Rigby, DS, Mel, me2} to test the accuracy of this decomposition. 

\renewcommand{\thefootnote}{\arabic{footnote}}

\section{Samples and Data Analysis}
The selection details for the [OIII] and 12$\mu$m samples are presented in \citet{me1} and \citet{12m} and are therefore only briefly mentioned here. The [OIII] sample was selected from SDSS Data Release 4 \citep{DR4}. These sources live in the Seyfert 2 locus of the BPT diagram \citep{BPT} according to the \citet{Kewley} demarcation, and the sample is complete to a flux limit of 4$\times 10^{-14}$ erg cm$^{-2}$ s$^{-1}$, totaling 20 sources. Seventeen of these Sy2s were observed with {\it XMM-Newton}. The 12$\mu$m sample is comprised of Sy2s from the {\it IRAS} point source catalog, is complete to a flux density limity of 0.3 Jy at 12$\mu$m and totals 31 galaxies. Twenty-eight of these sources have archival {\it Chandra} and/or {\it XMM-Newton} data.

In \citet{me1} and \citet{me3} we discussed in detail the X-ray data reduction, using XAssist \citep{xassist}, and broad-band X-ray (0.5 - 10 keV) spectral fitting of these sources with XSpec. Each galaxy was fit with an absorbed power law or when necessary, double absorbed power law, with the spectral indices initially tied together, simulating a partial covering geometry where a certain percentage of the transmitted X-ray emission is absorbed and the remainder scattered into the line of sight; Fe K$\alpha$ emission, when present, was modeled with a Gaussian component. To accomodate the possible presence of starburst emission in soft X-rays (0.5-2 keV), we included a thermal model (APEC) in these spectral fits which describes emission from a collisionally ionized plasma. Here we revisit more detailed modeling of the soft emission.

Since our aim is to study host galaxy star formation in tandem with AGN activity, we limit our analysis to the Sy2s where addition of a thermal model component results in a detection of thermal emission, rather than an upper limit. As a result, NGC 291, CGCG 064-017, CGCG 218-007 and 2MASX J13463217+6423247 were dropped from the [OIII] sample. We also exclude the three 12$\mu$m sources that were not detected in 2 - 10 keV X-rays: F08572+3915, NGC 5953 and NGC 7590. NGC 1068, part of the original 12$\mu$m sample, has soft X-ray emission dominated by AGN photoionization \citep[e.g.][]{N1068} and has therefore been omitted from this analysis. In total, 38 Seyfert 2 galaxies remain, 14 from the [OIII] sample and 24 from the 12$\mu$m sample.

As noted in \citet{me3}, several 12$\mu$m Sy2s had both {\it Chandra} and {\it XMM-Newton} archival observations. We tested if differences in the spectra for an individual source between the two observatories were due to aperture effects, since {\it Chandra} has a higher spatial resolution than {\it XMM-Newton}, or variability. Our aim in that analysis was to constrain the properties of the AGN, so we focused on the {\it Chandra} data to isolate the point source, using {\it XMM-Newton} information to help constrain the spectral fit for non-variable sources. In this analysis, we favor larger aperture extraction areas as they encompass extended X-ray emission due to host galaxy star formation. For the Sy2s in the 12$\mu$m sample with both {\it XMM-Newton} and {\it Chandra} observations, the {\it Chandra} extraction area was adjusted to match the area from {\it XMM-Newton} or to encompass all visible extended X-ray emission. For the sources with just {\it Chandra} data, we increased the extraction area from the analysis presented in \citet{me3} to ensure all extended X-ray emission is covered. In Tables \ref{o3_ap} and \ref{12m_ap}, we list the aperture sizes used to extract spectra for the [OIII] and 12$\mu$m Sy2s, respectively.

The spectral fitting results for the [OIII] and 12$\mu$m samples are summarized in Tables \ref{o3_apec} and \ref{12m_apec}. The broad-band X-ray spectra (0.5 - 10 keV) were fitted to most fully constrain the AGN contribution to the soft X-ray emission. Multiple spectra of the same source (from the three {\it XMM-Newton} detectors and/or multiple observations), were initially fit simultaneously with a multiplicative constant allowed to vary to account for differences in detector sensitivity. For variable sources, individual model parameters were fit independently. We imposed a lower column density (N$_{H}$) limit of that due to our Galaxy along the line of sight to the source. In some cases, the best fit absorption pegged at this lower limit and N$_{H}$ was subsequently frozen to this value. For the sources better described by a double power law model, which as noted above represents a partial covering geometry, the photon indices were tied together, with the normalizations left free to vary and an independent absorption attenuating the second power law component (N$_{H,2}$); the Sy2s best accommodated with a single power law can be identified by the blank entry in the N$_{H,2}$ column of Tables \ref{o3_apec} and \ref{12m_apec}. The lower temperature limit (kT) was set at 0.1 keV. Ten Sy2 galaxies were well fit by a second APEC component, significant at greater than the 3$\sigma$ level according to the f-test: 2MASX J08244333+2959238, NGC 424, NGC 1320, NGC 1386, NGC 4388, M-3-34-64, NGC 5194, Mrk 463, NGC 7582 and NGC 7674. This result indicates the presence of a two temperature gas, similar to the results of X-ray observations of starburst galaxies \citep{Dahlem, Strickland}: one phase at kT between 0.6-0.75 keV and a lower temperature gas at $\sim$0.15 keV, with the exception of M-3-34-64 which has two hot gas components at 0.77$^{+0.02}_{-0.02}$ keV and 2.85$^{+0.34}_{-0.34}$. 

We list the observed soft fluxes in Tables \ref{o3_flux} and \ref{12m_flux}: total 0.5-2 keV flux (F$_{0.5-2keV}$), power law component of soft X-ray flux (F$_{pow}$) and APEC component of soft X-ray flux (F$_{APEC}$). In the subsequent analysis, we omit the Sy2s with a poorly costrained APEC flux, i.e. the sources where the error is an order of magnitude or more higher or lower than the measured value (2MASX J11110693+0228477 and CGCG 242-028), leaving us with 35 Sy2s.

\section{Effects of X-ray Binaries}
Before assigning ``thermal'' (F$_{APEC}$) flux to starburst activity and power law emission (F$_{pow}$) to AGN, we need to account for the effects of X-ray binaries which contribute to the non-thermal emission. As noted above, high mass X-ray binaries indicate instantaneous star formation in the host galaxy and this effect is incorporated into X-ray SFR calibrations. Since X-ray binaries boost the non-thermal flux, F$_{pow}$ will overestimate the AGN emission while F$_{APEC}$ under-represents host galaxy star formation. To assess the contribution of X-ray binaries to the observable emission, we model just the extended emission when the AGN is resolvable. For the remaining sources, we apply a correction to F$_{APEC}$ and F$_{pow}$ using results of X-ray analysis of star-forming galaxies from the XSINGS sample (Ptak et al. 2012).

Spatially resolved emission from the narrow line region (NLR) can also potentially affect the results, over-predicting the luminosity attributable to host galaxy star formation. For a handful of 12$\mu$m sources, the major radius ($R_{maj}$) of the NLR was measured from \citet{schmitta}, while we used the relations from \citet{schmittb} to estimate NLR $R_{maj}$ values for the remaining 12$\mu$m sources and the [OIII] sample. For the [OIII] Sy2s, the NLR radius ranges from 280 - 900 pc, with a median size of $\sim$0.5 kpc. Similarly, the 12$\mu$m selected Sy2s have NLR sizes ranging from 80 - 1320 pc, with a median $R_{maj}$ value of $\sim$0.39 kpc. For all [OIII] and a majority of 12$\mu$m sources, the NLR is an order of magnitude or more smaller than the X-ray extraction radius. The NLR is thus likely not resolvable at these scales, so we only focus on quantifying the effects of X-ray binaries on the observed emission. One notable exception is IC 5063, where the radius of the NLR ($\sim$1.3 kpc) comprises a signficant fraction of the X-ray extraction radius (3 kpc). We will return to this Sy2 in Section \ref{results}. Another potential bias could be due to photoionized gas from the NLR being indistinguishable from collisionally ionized gas at low CCD resolution. However as we show in Section \ref{results}, such an effect does not systematically affect our results.

\subsection{{\it Chandra} Imaging}

Due to {\it Chandra}'s arc second resolution and the close proximity of the 12$\mu$m Sy2s, the central point source (i.e. AGN) is resolvable for a subset of our sample. To accurately constrain the AGN position, we ran wavedetect on 2-10 keV {\it Chandra} images of the 12$\mu$m sample. Spectra were then extracted from the remaining emission, omitting the AGN identified by wavedetect. Similar to above, the broad band X-ray emission was fitted with (an) absorbed power law(s) and an APEC component. We only include those sources where the spectra from the residual emission have a high enough signal-to-noise to be fit with the $\chi^2$ statistic (minimum grouping of 10 counts per bin), therefore dropping the sources where the X-ray emission is concentrated in the nucleus: NGC 424, TOLOLO 1238-364, F05189-2524 and NGC 5347. The summary of the spectral fitting for these 11 sources is listed in Table \ref{pt_src_rem}, and the associated total soft X-ray flux from the unresoloved emission (i.e. host galaxy emission after AGN removal) is listed in the second column of Table \ref{12m_flux}. To first order, the total soft X-ray flux from the unresolved emission can be considered an estimate of host galaxy star formation, L$_{x,SF}$.  L$_{x,AGN}$ is then estimated by taking the difference between L$_{x,SF}$ and L$_{0.5-2keV}$ for these Sy2s, with the associated error derived by scaling the error of total soft X-ray luminosity by the ratio of the AGN luminosity to the total luminosity. 

\subsection{\label{xsings} Star-forming Galaxy Template}

We have used the results of XSINGS to estimate the contribution of X-ray binaries (XRBs) to the soft X-ray emission in host galaxies. The XSINGS project entails the analysis of 96 Chandra observations of 56 Spitzer Infrared Nearby Galaxy Survey (SINGS) galaxies (Kennicutt et al, 2003).  SINGS is designed to cover a wide range of galaxy properties such as morphological type, star formation rate (SFR), stellar mass and metallicity, and the main aim of XSINGS is to study the relationships between X-ray emission (X-ray binaries, nuclear activity and hot ISM) with these properties.  In Ptak et al. (2012, in prep.) the spatially-averaged X-ray emission of XSINGS galaxies is discussed and compared to the summed X-ray binary luminosities for galaxies with significant detections (X-ray binary catalogs for all XSINGS galaxies will be presented in Jenkins et al. 2013).
 XRBs emit as a power law, but provide a measure of the instantaneous SFR along with thermal emission in soft X-rays. We therefore have:
\begin{equation}
L_{x,SF} = L_{APEC} + L_{XRB}
\end{equation}
\begin{equation}
L_{x,AGN} = L_{pow} - L_{XRB}
\end{equation}

We further assume that a certain fraction of X-ray emission from XRBs contributes to host galaxy star formation:
\begin{equation}
L_{XRB} = R \times L_{x,SF}
\end{equation}

Using the 22 pure star-forming galaxies in XSINGS (i.e. non Sy2s and non LINERs), we have calculated the mean ratio of X-ray emission from resolved non-nuclear point sources (i.e. X-ray binaries) to total X-ray emission in the 0.5-2 keV band ($R$), deriving $R = 0.51 \pm 0.26$. Solving the above equations, we obtain:
\begin{equation}
L_{x,SF} = \frac{L_{APEC}}{1-R}
\end{equation}
\begin{equation}
L_{x,AGN} = L_{pow} - R\frac{L_{APEC}}{1-R}
\end{equation}

To account for uncertainties in the luminosities derived from spectral fitting (i.e. L$_{APEC}$ and L$_{pow}$) and the wide spread in $R$ values, we have performed Monte Carlo simulations to derive L$_{x,SF}$ and L$_{x,AGN}$. We have drawn 1000 random values for L$_{APEC}$, L$_{pow}$ and $R$ from a Gaussian distribution centered on the best-fit values of L$_{APEC}$ and L$_{pow}$ (and mean value of $R$), with $\sigma$ corresponding to the errors derived from spectral fitting (and standard deviation on $R$). Using these simulated parameters and Eqs. 4 and 5, we calculated 1000 simulated values for L$_{x,SF}$ and L$_{x,AGN}$.  The mode of these distributions is then taken as L$_{x,SF}$ and L$_{x,AGN}$, with the 68\% confidence interval as the associated errors.

For the subset of Sy2s with high signal-to-noise {\it Chandra} spectra of the soft unresolved emission, we compare the results of L$_{x,SF}$ (L$_{x,AGN}$) derived via {\it Chandra} imaging analysis and the simulation approach described above. Here, L$_{x,SF,Chandra}$ (L$_{x,AGN,Chandra}$) represents the soft X-ray luminosity due to star formation (AGN) based on spectral fitting of just the unresolved emission, while L$_{x,SF,XSINGS}$ (L$_{x,AGN,XSINGS}$) is the estimate of the star formation (AGN) component of the soft X-ray flux calculated via simulations.

\begin{figure}[ht]
\centering
\subfigure[]{\includegraphics[scale=0.35,angle=90]{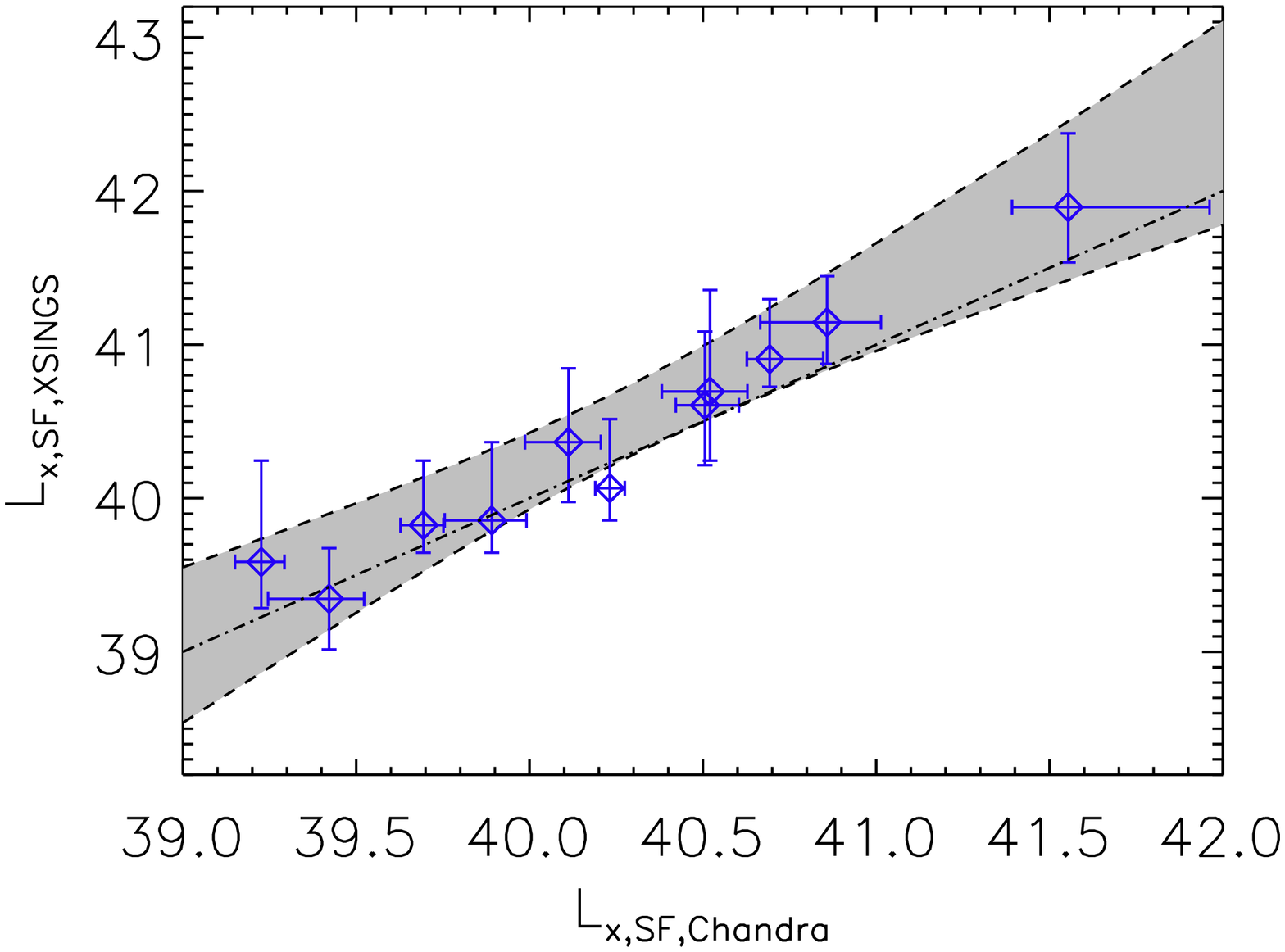}}
\subfigure[]{\includegraphics[scale=0.35,angle=90]{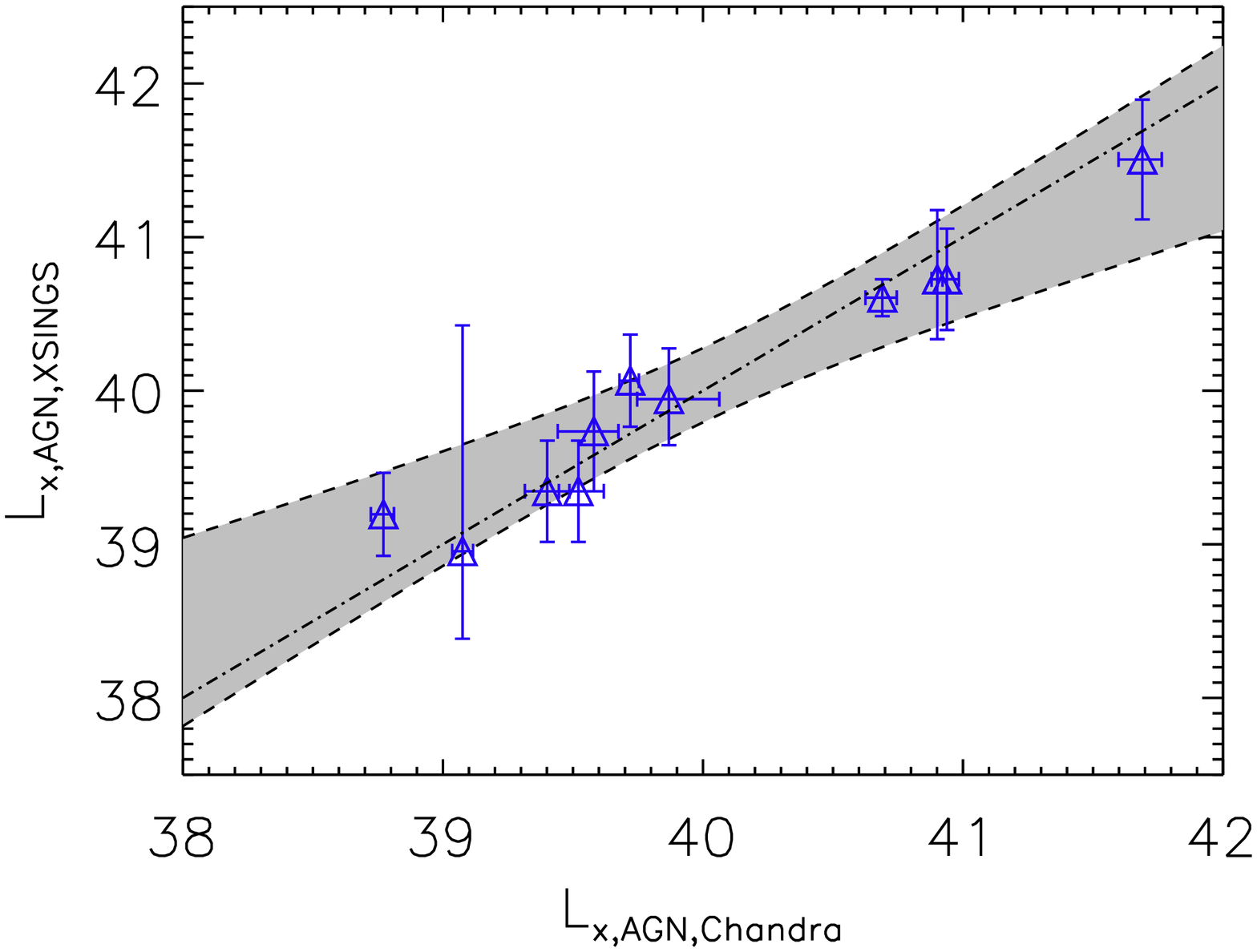}}
\caption[]{\label{chandra_v_sings} Comparison of using {\it Chandra} observations to decompose the soft X-ray emission into star formation and AGN components (L$_{x,SF,Chandra}$ and L$_{x,AGN,Chandra}$, respectively) with estimates of starburst and AGN activity calculated via simulations using XSINGS data of normal galaxies and luminosities derived from spectral fitting (i.e., L$_{x,SF,XSINGS}$, L$_{x,AGN,XSINGS}$). The grey shaded regions enclosed by dashed lines indicate the 3$\sigma$ confidence interval from a Bayesian linear regression fit \citep{Kelly}. The dotted-dashed lines denote where the two quantities are equal. These two methods of estimating the star formation and AGN contributions to the soft X-ray emission are consistent.}
\end{figure}

To account for errors in both parameters, we use a Bayesian approach to linear regression \citep{Kelly}. Both here and in Section \ref{results}, the high and low error bars on the X-ray flux were averaged to provide symmetric errors for the linear regression routine. We plot L$_{x,SF,XSINGS}$ as a function of L$_{x,SF,Chandra}$ and L$_{x,AGN,XSINGS}$ against L$_{x,AGN,Chandra}$ in Figure \ref{chandra_v_sings}. The grey shaded regions enclosed by the dashed lines delineate the 3$\sigma$ confidence level on the regression line, based on the median of the posterior distribution of the slope, intercept and mean distribution of the independent variable, with the dotted-dashed line denoting the line of equality. Though the L$_{x,SF,XSINGS}$ values are somewhat systematically higher than L$_{x,SF,Chandra}$, they do agree within the 3$\sigma$ contours. The agreeement among these independent methods suggests that using XSINGS galaxies as a template to correct the L$_{APEC}$ and L$_{pow}$ values for the presence of X-ray binaries provides a reasonable estimate to decompose the soft X-ray emission in the absence of resolveable nuclear emission.

In the subsequent analysis (i.e. Section \ref{results}), we use L$_{x,SF,Chandra}$ and L$_{x,AGN,Chandra}$ as L$_{x,SF}$ and L$_{x,AGN}$ for the 11 {\it Chandra} sources with high signal-to-noise. L$_{x,SF}$ and L$_{x,AGN}$ for the remaining 24 Sy2s are estimated via the Monte Carlo simulations outlined above. These values are listed in Tables \ref{o3_lum} and \ref{12m_lum} for the [OIII] and 12$\mu$m samples, respectively.

\section{Constraints from Infrared Data}
We compare our X-ray decomposition with information gleaned from high resolution infrared spectroscopy, using diagnostics that cleanly trace the separate processes of AGN and star formation activity. The flux of the [OIV]26$\mu$m line has been shown to accurately describe intrinsic AGN flux \citep{Rigby, DS, Mel, me2} while the [NeII]12.8$\mu$m emission reliably tracks star formation in both quiescent and active systems (LaMassa et al., 2012). We therefore use these fluxes to decompose the soft X-ray emission into a star forming and an AGN component by solving L$_{0.5-2keV}$ = $\alpha \times$L$_{[NeII]}$ + $\beta \times$L$_{[OIV]}$ using ordinary least squares multiple linear regression on both the 12$\mu$m and [OIII] samples, obtaining $\alpha$=1.07 and $\beta$=0.16 \citep[see][for calculated values of L$_{[OIV]}$ and L$_{[NeII]}$ for these samples]{me2, me4}. Here L$_{0.5-2keV}$ represents the total soft X-ray emission (i.e. column 1 in Tables \ref{o3_flux} and \ref{12m_flux}). 

\begin{figure}[ht]
\centering
{\includegraphics[scale=0.5,angle=90]{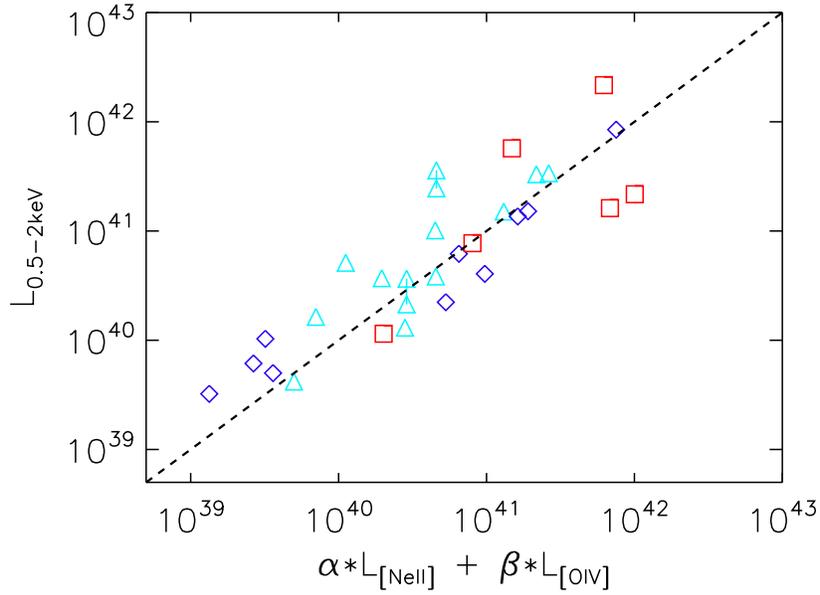}}
\caption[]{\label{softx_ne2_o4} The results of decomposing the soft X-ray flux (0.5-2 keV) into a star formation and AGN component, using the luminosity of the [NeII] line as a proxy for the former and the luminosity of the [OIV] line to parameterize the latter. The constants $\alpha$ and $\beta$ were calculated using ordinary least squares multiple linear regression, where we derive $\alpha$=1.07 and $\beta$=0.16. The overplotted line indicates where the two quantities are equal. Cyan triangles represent the 12$\mu$m Sy2s (with the X-ray variable sources connected by a vertical line), blue diamonds denote the subset of 12$\mu$m sources with high signal-to-noise {\it Chandra} imaging of the unresolved emission and red squares mark the [OIII] selected sources.}
\end{figure}

Figure \ref{softx_ne2_o4} illustrates the results of this decomposition. We consider the X-ray variable Sy2s, NGC 5506 and NGC 7172 as independent data points, but we do average the soft X-ray flux for the latter two NGC 5506 observations, which are consistent. Though we have only one IR measurement for each of these sources, we do not expect the infrared emission to vary as much as the X-ray emission. We color code the sources in Figure \ref{softx_ne2_o4} according to sample: cyan triangles represent the 12$\mu$m Sy2s (with the X-ray variable sources connected by a vertical line), blue diamonds denote the subset of 12$\mu$m sources with high signal-to-noise {\it Chandra} imaging of the unresolved emission and red squares mark the [OIII] selected sources. The overplotted line indicates where the two quantities are equal. With the exception of a handful of outliers, a relatively good agreement is present between the observed soft X-ray emission and the IR decomposition. We note that this consistency holds for both the 12$\mu$m and [OIII] samples, with no apparent systematic offset introduced from different sample selection techniques. In the following analysis, we therefore use $\alpha \times$L$_{[NeII]}$ and $\beta \times$L$_{[OIV]}$ as an independent estimate of the starburst and AGN contribution, respectively, to the soft X-ray emission.

\section{\label{results} Results: Comparison of Soft X-ray Decomposition with IR}

\begin{figure}[ht]
\centering
\subfigure[]{\includegraphics[scale=0.40,angle=90]{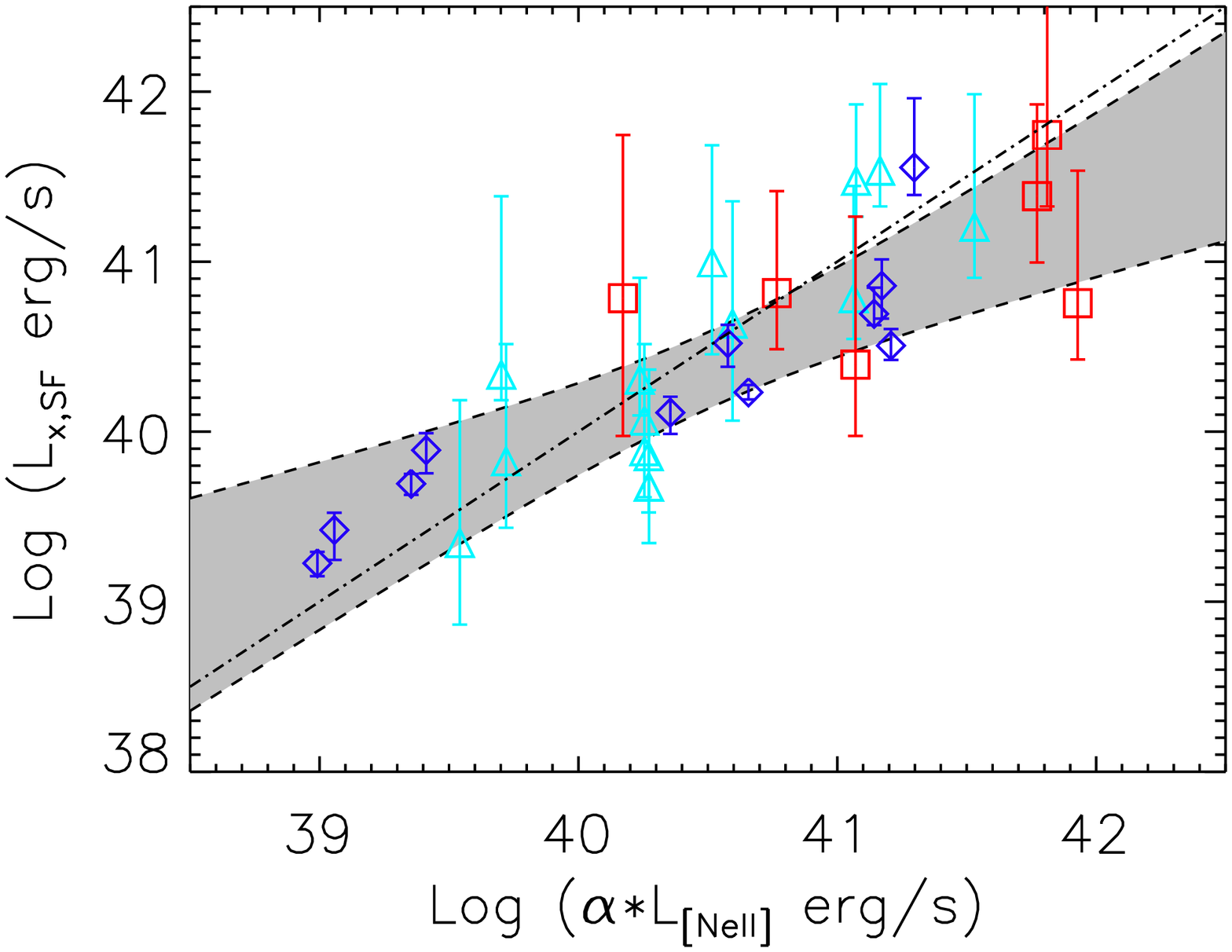}}
\subfigure[]{\includegraphics[scale=0.40,angle=90]{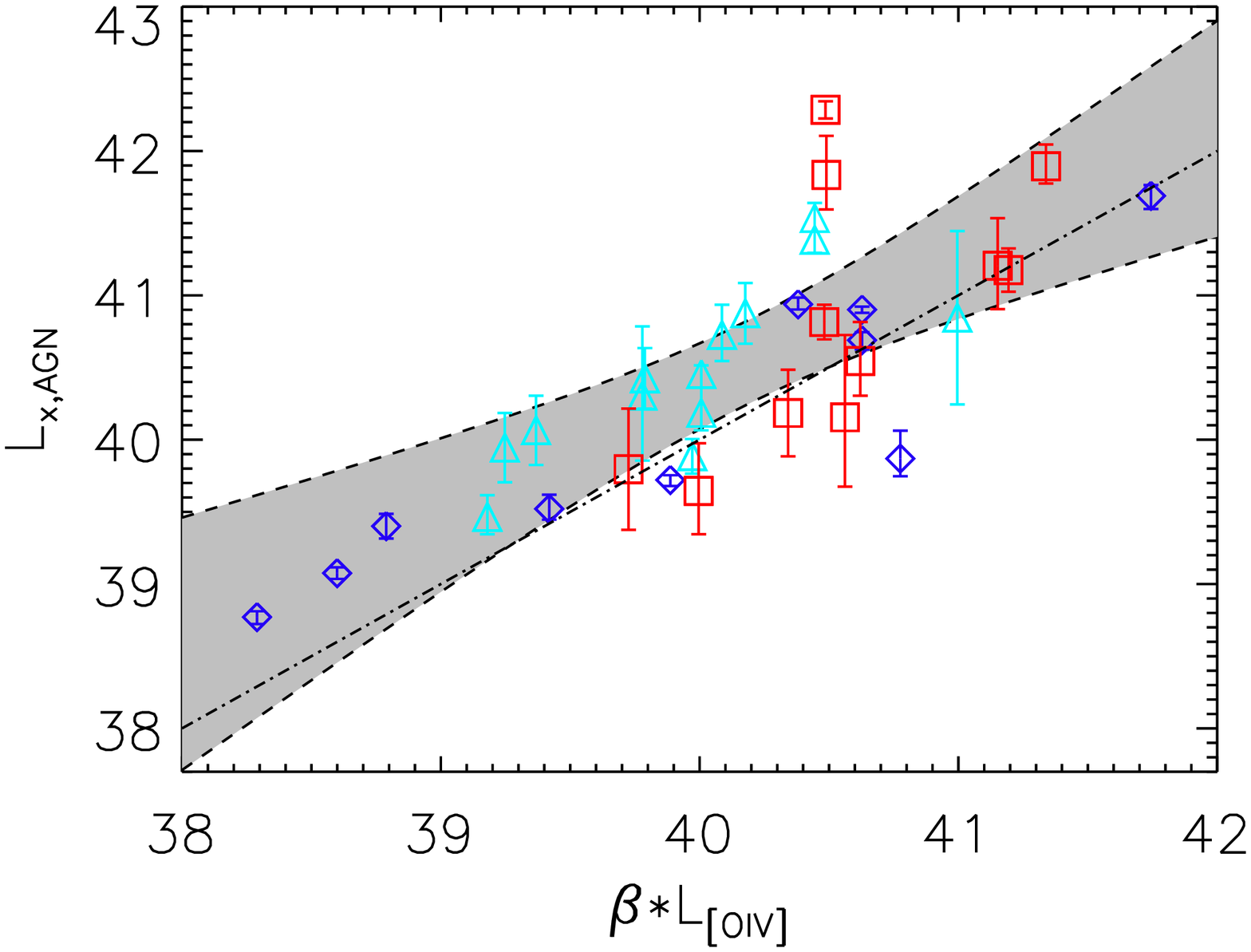}}
\caption[]{\label{lx_ir}Left: L$_{x,SF}$ vs $\alpha \times$L$_{[NeII]}$. Right: L$_{x,AGN}$ vs $\beta \times$L$_{[OIV]}$. In both plots, the dashed-dotted line indicates the line of equality, while the gray shaded regions illustrate the 3$\sigma$ confidence interval from Bayesian linear regression. Color coding is the same as Figure \ref{softx_ne2_o4}. L$_{x,SF}$ and L$_{x,AGN}$ are derived from spectral fitting of the unresolved {\it Chandra} emission for the blue diamond data points while these parameters are estimated by scaling L$_{APEC}$ and L$_{pow}$ by as noted in the text. The soft X-ray and IR decomposition approximately agree at the 3$\sigma$ level, with small deviations appearing at luminosities above 10$^{41}$ erg/s/cm$^2$ for the star formation parameterization and below 2$\times 10^{41}$ erg/s/cm$^2$ for the AGN decomposition.}
\end{figure}

We plot L$_{x,SF}$ as a function of $\alpha \times$ L$_{[NeII]}$ and L$_{x,AGN}$ as a function of $\beta \times$ L$_{[OIV]}$ in Figure \ref{lx_ir}. We note that the abscissa errors are on the order of the symbol size, due to the 5$\sigma$ detection limit we imposed on IR detections (see LaMassa et al. 2012), and therefore are not plotted in Figure \ref{lx_ir}, though they are included in the errors for the Bayesian linear regression routine. Due to the large error on L$_{x,AGN}$ for NGC 7674, this source was dropped from the L$_{x,AGN}$ vs. $\beta \times$ L$_{[OIV]}$ analysis.

In Figure \ref{lx_ir}, a dotted-dashed line is plotted to indicate where the two quantities are equal while the grey shaded region illustrates the 3$\sigma$ confidence interval on the regression line. L$_{x,SF}$ and L$_{x,AGN}$ are approximately equivalent to $\alpha \times$ L$_{[NeII]}$ and $\beta \times$ L$_{[OIV]}$, respectively, albeit with wide scatter for a handful of individual sources. Deviations of the regression fit from equality at greater than the 3$\sigma$ level are slight, and appear at higher luminosities for the star formation decomposition ($> 10^{41}$ erg/s/cm$^2$) and moderate luminosities for the AGN parameterization ($10^{39} \lesssim$ L$_{x,AGN} \lesssim 5\times 10^{41}$ erg/s/cm$^2$). As mentioned previously, the NLR for IC 5063 takes up a signficant fraction of the the X-ray extraction region, where contributions from resolved  NLR emission could potentially bias the X-ray estimate of host galaxy star formation. However, as seen in Figure \ref{lx_ir}, where IC 5068 is the blue diamond at $\alpha \times$ L$_{[NeII]}$ = 40.4 dex and L$_{x,SF}$ = 40.1 dex, this source lies well within the shaded region. Hence NLR contamination appears to be negligible, possibly due to the removal of the AGN from the {\it Chandra} imaging analysis.

We note that aperture bias can contribute to these slight disagreements: [NeII] ([OIV]) is measured through the Short-High (Long-High) module on the IRS spectrograph on {\em Spitzer}, corresponding to sizes 4.7''$\times$11.3'' (11.1''$\times$22.3''), whereas the X-ray apertures capture the full emission from the host galaxy (see Tables \ref{o3_ap} and \ref{12m_ap}). Hence, biases can be introduced from comparing parameters that sample different physical scales of the host galaxy, introducing scatter into the relation. In the AGN parameterization, there are 4 obvious outliers from the relation: IC 0486 and Mrk 463 (members of the [OIII] sample with over-predicted L$_{x,AGN}$, red squares in Figure \ref{lx_ir}), NGC 4388 (from the 12$\mu$m sample with under-predicted L$_{x,AGN}$, blue diamond in Figure \ref{lx_ir}) and NGC 5506 (from the 12$\mu$m sample with over-predicted L$_{x,AGN}$, an X-ray variable source with a vertical line connecting the cyan triangles in \ref{lx_ir}). Interestingly, there are no correlations between outliers in the L$_{x,AGN}$ parameterization and physical size of X-ray emitting region (where mis-match beween aperture sizes between {\it Chandra} or {\it XMM-Newton} and {\it Spitzer} could introduce disagreements), quality of soft X-ray spectra, sample selection, source variability or even method for estimating L$_{x,AGN}$. In general, however, deviations of the best-fit trends between L$_{x,SF}$ versus $\alpha \times$ L$_{[NeII]}$ and L$_{x,AGN}$ versus $\beta \times$ L$_{[OIV]}$ from equality at the 3$\sigma$ level are slight. Such an agreement demonstrates that our method of using low resolution CCD spectroscopy to decompose the soft X-ray emission into a star-forming and an AGN component is moderately successful, with greater consistency in the star formation estimation. L$_{x,SF}$ can then be used to estimate the host galaxy SFR in Sy2s using the L$_{0.5-2keV}$ calibrations from \citet{Ranalli} and \citet{PS}.

\subsection{Comparison with Far-Infrared Derived SFRs}
For the 12$\mu$m sample, we have tested the efficacy of our decomposition by deriving a soft X-ray SFR using the \citet{PS} calibration and comparing this with the far-infrared (FIR) SFR using the \citet{Kennicutt} calibration, using IRAS flux densities reported in \citet{12m}. The [OIII] sample, however, lacks a reliable independent host-galaxy wide SFR. These sources were not detected by {\it IRAS} and the SDSS SFRs derived through the 3'' fiber \citep[see][for a discussion]{me4} only covers the innermost regions of the galaxy, while the X-ray aperture captures the full emission. We therefore use only the 12$\mu$m sample to test the X-ray derived SFRs.

\begin{figure}[ht]
\centering
{\includegraphics[scale=0.5,angle=90]{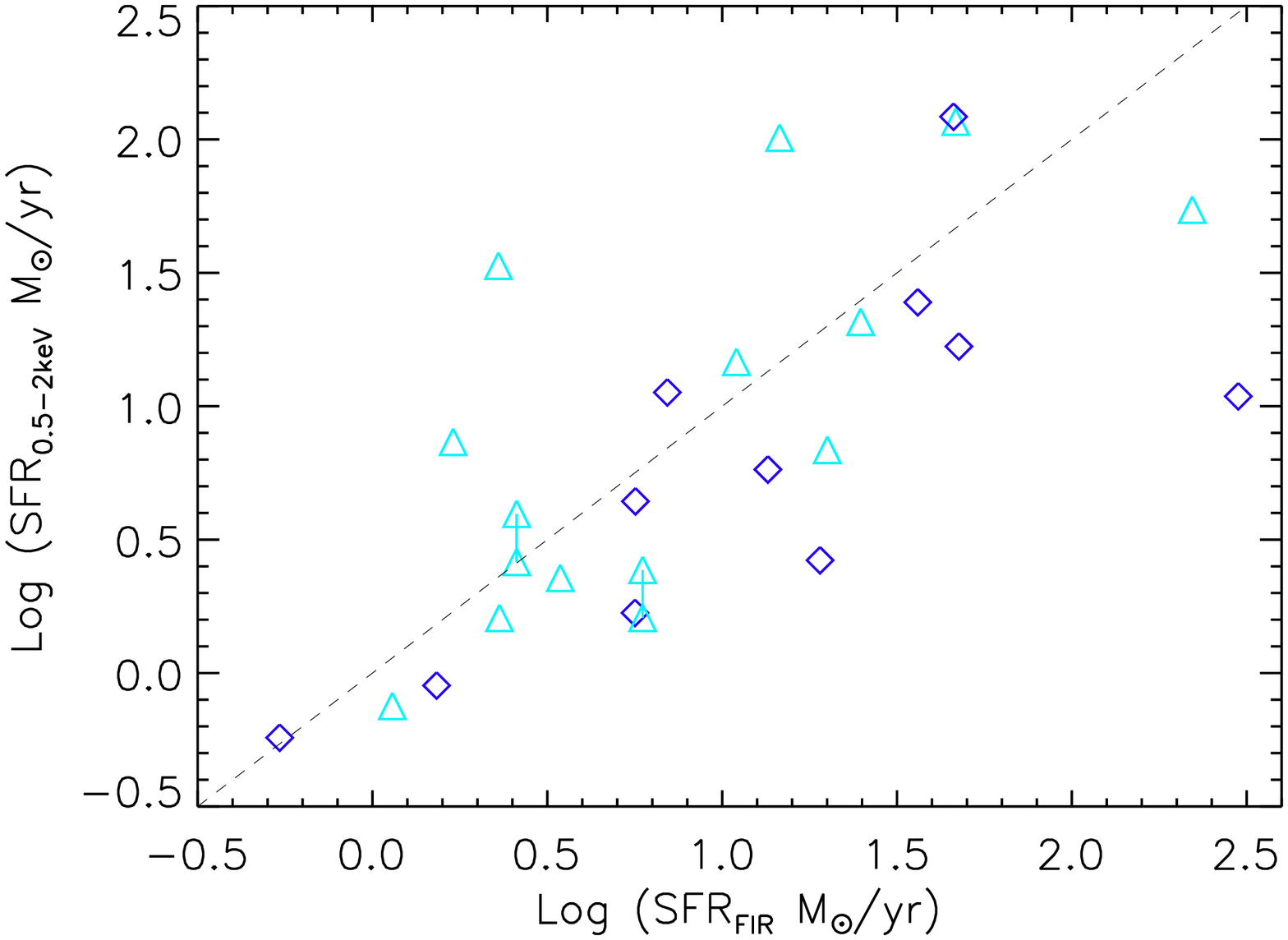}}
\caption[]{\label{firsfr_xsfr} X-ray derived SFRs (SFR$_{0.5-2keV}$) using our estimate of L$_{x,SF}$ and the calibration of \citet{PS} as a function of FIR derived SFRs (SFR$_{FIR}$), using {\it IRAS} 60$\mu$m and 100$\mu$m luminosities and \citet{Kennicutt}'s calibration, for the 12$\mu$m sample. The overplotted dashed line shows where the two SFRs are equal. A relatively good agreement exists, with the two most significant outliers with the largest SFR$_{FIR}$ values (Arp 220 and F05189-2524) being ULIRGs.}
\end{figure}

To estimate SFR$_{FIR}$, we used the sum of 60$\mu$m and 100$\mu$m {\it IRAS} flux densities. We note that the \citet{Kennicutt} calibration is based on MIR through FIR luminosities for starburst galaxies, and in such systems, including the IRAS 12$\mu$m and 25$\mu$m luminosities as part of the FIR luminosity is important. However, these MIR luminosities have significant contributions from dust heated by the AGN, so we only consider the 60$\mu$m and 100$\mu$m luminosities when deriving SFR$_{FIR}$, which are less contaminated by the AGN. As shown in Figure \ref{firsfr_xsfr}, a good agreement exists between SFR$_{0.5-2keV}$ and SFR$_{FIR}$, despite the latter SFR parameterization being most appropriate for starburst galaxies. This agreement suggests that any unresolved photoionized emission from the NLR does not introduce any systematic offsets into our results, but rather operates at a level less than the observed scatter. The 2 Sy2s that are the greatest outliers (i.e., largest SFR$_{FIR}$ values) are Arp 220 and F05189-2524, which are Ultra Luminous Infrared Galaxies (ULIRGs). 

\section{Conclusion}
We have modeled the 0.5-10 keV spectra of two homogeneous samples of Seyfert 2 galaxies to disentangle the starburst and AGN emission in soft X-rays (0.5-2 keV). Eleven of these Sy2s, observed with {\it Chandra}, had high signal-to-noise unresolved emission after removing the AGN. We derive L$_{x,SF}$ from spectrally fitting this unresolved emission and assign L$_{x,AGN}$ to the difference between the total soft X-ray emission and L$_{x,SF}$.

The remaining 24 sources were decomposed into an AGN and star formation component by modeling the soft X-ray emission with a power law and thermal model. The luminosities of these sources were converted to L$_{x,SF}$ and L$_{x,AGN}$ values based on a scaling factor derived from XSINGS analysis on normal star forming galaxies. To account for errors on the luminosities derived from spectral fits and the scaling factor, we executed Monte Carlo simulations, assuming a Gaussian distribution of random variables centered on the best-fit values of L$_{APEC}$, L$_{pow}$ and $R$, with the spread corresponding to the errors on these parameters. Using Eqs. 4 and 5, we calculate L$_{x,SF}$ and L$_{x,AGN}$ from the 1000 simulated parameters and use the mode of this distribution as our estimate of the X-ray luminosities due to star formation and AGN activity.

Our conclusions are summarized as follows:

\begin{itemize}

\item The soft X-ray spectra of 10 Sy2s were well fitted by two thermal model components, indicating the presence of a two temperature gas. This result is similar to what has been observed in starbust galaxies \citep{Dahlem, Strickland}.

\item For the subset of 11 Sy2s with high signal-to-noise {\it Chandra} imaging of unresolved host galaxy emission, estimates of the soft X-ray emission due to star formation and AGN from both {\it Chandra} imaging analysis and estimates from the Monte Carlo simulations agree at the 3$\sigma$ level. In the absence of resolved nuclear emission, scaling L$_{pow}$ and L$_{AGN}$ by the factors derived from XSINGS analysis is thus a reasonable method to estimate the AGN and star formation contributions to the soft X-ray emission.

\item The independent decompositions of the soft X-ray luminosity into a star formation and AGN component using IR data as a proxy and scaling L$_{APEC}$ and L$_{pow}$ based on XSINGS galaxies largely agree within a 3$\sigma$ confidence interval. Deviations of the best fit regression line from this agreement are slight and appear at higher luminosities ($>10^{41}$ erg/s/cm$^2$) in the star formation decomposition and at moderate luminosities ($10^{39} \lesssim$ L$\lesssim 5\times 10^{41}$ erg/s/cm$^2$) when describing AGN emission. Though more scatter in individual sources is evident in the AGN decomposition, the star formation relationship is more consistent among both methods.

\item Comparison of our calculated X-ray SFR using the \citet{PS} calibration with an FIR derived SFR from \citet{Kennicutt} for the 12$\mu$m sample shows general agreement.

\end{itemize}

We have demonstrated that analysis of low resolution CCD X-ray spectra can effectively disentangle emission from AGN activity and star formation in 0.5-2keV X-rays. Using the decomposition we have presented, L$_{x,SF}$ can be used to cleanly estimate the SFR in Sy2s using existing calibrations \citep[e.g.]{Ranalli,PS}, complementing previous studies in the optical and infrared and providing a more panchromatic view of the interplay between SMBH accretion and star formation.

\clearpage
\begin{landscape}
\begin{deluxetable}{lccc}

\tablewidth{0pt}
\tablecaption{\label{o3_ap}[OIII] Sample: Aperture Extraction Areas\tablenotemark{1}}
\tablehead{

\colhead{Galaxy} & \colhead{$z$} & \colhead{Aperture Radius ('')}    & \colhead{Aperture Radius (kpc)\tablenotemark{2}} \\
                 &               & \colhead{PN/MOS1/MOS2} & \colhead{PN/MOS1/MOS2} }

\startdata

Mrk 0609                & 0.034 & 65/34/35 & 44/23/24 \\

IC 0486                 & 0.027 & 34/33/35 & 19/18/19 \\

2MASX J08035923+2345201 & 0.029 & 17/20/18 & 10/12/11 \\

2MASX J08244333+2959238 & 0.025 & 40/40/40 & 21/21/21 \\

2MASX J10181928+3722419 & 0.049 & 20/22/22 & 20/22/22 \\

2MASX J11110693+0228477\tablenotemark{3} & 0.035 & 18/-/-  & 13 \\

CGCG 242-028            & 0.026 & 19/22/13 & 10/12/7  \\

SBS 1133+572            & 0.050 & 20/16/17 & 20/16/17 \\

Mrk 1457                & 0.049 & 19/13/13 & 18/13/13 \\

2MASX J11570483+5249036 & 0.036 & 35/35/35 & 26/26/26 \\

2MASX J12183945+4706275\tablenotemark{4} & 0.094 & -/30/31 & 53/55 \\

2MASX J12384342+0927362 & 0.083 & 33/30/30 & 52/47/47 \\

NGC 5695                & 0.014 & 31/22/17 & 9/7/5    \\

\enddata

\tablenotetext{1}{[OIII] sources were observed with {\it XMM-Newton} only. PN, MOS1 and MOS2 refer to the three detectors onboard {\it XMM-Newton}.}
\tablenotetext{2}{Conversion from arc seconds to kpc is based on the cosmology of H$_0$=70 km/s/Mpc, $\Omega_M$=0.27 and $\Omega_\Lambda$ = 0.73.}
\tablenotetext{3}{Only PN spectrum fit since MOS1 and MOS2 spectra suffered from low signal-to-noise.}
\tablenotetext{4}{Source fell on chip gap in PN detector so only spectra from MOS1 and MOS2 were fit.}
\end{deluxetable}

\end{landscape}

\clearpage

\begin{landscape}
\begin{deluxetable}{lccccc}
\small
\tablewidth{0pt}
\tablecaption{\label{12m_ap}12$\mu$m Sample: Aperture Extraction Areas\tablenotemark{1}}
\tablehead{
\colhead{Galaxy} & \colhead{$z$} & \colhead{Observatory} & \colhead{ObsID} & \colhead{Aperture Radius('')} & \colhead{ApertureRadius (kpc)\tablenotemark{2}} \\
                 &                       &                 & \colhead{PN/MOS1/MOS2}      & \colhead{PN/MOS1/MOS2}}

\startdata

NGC 0424        & 0.012 & {\it XMM-Newton} & 00029242301 & 34/33/33 & 8/7/7    \\
                &       & {\it Chandra}    & 03146       & 15       & 3        \\

NGC 1144        & 0.028 & {\it XMM-Newton} & 0312190401  & 36/33/33 & 20/18/18 \\

NGC 1320        & 0.009 & {\it XMM-Newton} & 0405240201  & 37/35/35 & 6/6/6    \\

NGC 1386        & 0.003 & {\it XMM-Newton} & 0140950201  & 45/40/40 & 2/2/2    \\
                &       & {\it Chandra}    & 04076       & 19       & 1        \\

NGC 1667        & 0.015 & {\it XMM-Newton} & 0200660401  & 32/22/34 & 10/7/10  \\

F05189-2524     & 0.043 & {\it XMM-Newton} & 0085640101  & 41/31/33 & 35/26/28 \\
                &       & {\it Chandra}    & 02034       & 40       & 34       \\
                &       & {\it Chandra}    & 03432       & 40       & 34       \\

NGC 3982        & 0.004 & {\it XMM-Newton} & 0204651201  & 34/37/44 & 3/3/4    \\
                &       & {\it Chandra}    & 04845       & 34       & 3        \\

NGC 4388        & 0.008 & {\it XMM-Newton} & 0110930701  & 36/35/43 & 7/7/8    \\
                &       & {\it XMM-Newton} & 0110930301  & 37/32/33 & 7/6/6    \\
                &       & {\it Chandra}    & 01619       & 36       & 7        \\

NGC 4501        & 0.008 & {\it XMM-Newton} & 0112550801  & 18/22/33 & 3/4/6    \\
                &       & {\it Chandra}    & 02922       & 20       & 4        \\

TOLOLO 1238-364 & 0.011 & {\it Chandra}    & 04844       & 20        & 5       \\

NGC 4968        & 0.010 & {\it XMM-Newton} & 0002940101  & 17/18/14 & 4/4/3    \\
                &       & {\it XMM-Newton} & 0200660201  & 21/18/18 & 5/4/4    \\

M-3-34-64       & 0.017 & {\it XMM-Newton} & 0206580101  & 59/34/38 & 21/12/14 \\

NGC 5135        & 0.014 & {\it Chandra}    & 02187       & 20       & 6        \\

NGC 5194\tablenotemark{3} & 0.002 & {\it XMM-Newton} & 0112840201  & 65/42/49 & 3/2/2 \\
                          &       & {\it XMM-Newton} & 0303420101  & -/56/49  & -/2/2 \\
                          &       & {\it XMM-Newton} & 0303420201  & -/61/46  & -/3/2 \\
                          &       & {\it Chandra}    & 00354       & 40       & 2     \\
                          &       & {\it Chandra}    & 01622       & 40       & 2     \\
                          &       & {\it Chandra}    & 03932       & 40       & 2     \\ 

NGC 5347        & 0.008 & {\it Chandra}    & 04867       & 6        & 1        \\

Mrk 463         & 0.050 & {\it XMM-Newton} & 0094401201  & 49/45/45 & 49/45/45 \\  
                &       & {\it Chandra}    & 04913       & 40       & 40       \\

NGC 5506\tablenotemark{4} & 0.006  & {\it XMM-Newton} & 0013140101  & 60/-/-  & 9/-/-   \\
                          &        & {\it XMM-Newton} & 0013140201  & 54/-/-  & 8/-/-   \\
                          &        & {\it XMM-Newton} & 0201830201  & 56/-/53 & 8/-/8   \\
                          &        & {\it XMM-Newton} & 0201830301  & 52/-/64 & 8/-/9   \\
                          &        & {\it XMM-Newton} & 0201830401  & 55/-/55 & 8/-/8   \\
                          &        & {\it XMM-Newton} & 0201830501  & 68/-/98 & 10/-/14 \\
                          &        & {\it XMM-Newton} & 0554170201  & 64/-/-  & 9/-/-   \\
                          &        & {\it XMM-Newton} & 0554170101  & 66/-/-  & 10/-/-  \\

Arp 220         & 0.018  & {\it XMM-Newton} & 0101640801  & 20/21/25 & 8/8/9   \\
                &        & {\it XMM-Newton} & 0101640901  & 28/25/23 & 11/9/9  \\
                &        & {\it XMM-Newton} & 0205510201  & 34/28/22 & 13/11/8 \\
                &        & {\it Chandra}    & 00869       & 40       & 15      \\ 

NGC 6890        & 0.008  & {\it XMM-Newton} & 0301151001  & 17/13/15 & 3/2/2   \\

IC 5063         & 0.011  & {\it Chandra}    & 07878       & 12       & 3       \\

NGC 7130        & 0.016  & {\it Chandra}    & 02188       & 17       & 5       \\

NGC 7172        & 0.009  & {\it XMM-Newton} & 0147920601  & 52/41/34 & 8/7/5   \\
                &        & {\it XMM-Newton} & 0202860101  & 61/59/60 & 10/9/10 \\
                &        & {\it XMM-Newton} & 0414580101  & 53/59/55 & 8/9/9   \\

NGC 7582        & 0.005  & {\it XMM-Newton} & 0112310201  & 44/41/41 & 4/4/4   \\
                &        & {\it XMM-Newton} & 0204610101  & 60/43/47 & 6/4/4   \\
                &        & {\it Chandra}    & 00436       & 43       & 4       \\
                &        & {\it Chandra}    & 02319       & 43       & 4       \\

NGC 7674        & 0.029  & {\it XMM-Newton} & 0200660101  & 19/18/16 & 11/10/9 \\

\enddata

\tablenotetext{1}{For {\it XMM-Newton} observations, the extraction areas for the PN, MOS1 and MOS2 detectors are listed separately.}
\tablenotetext{2}{Conversion from arc seconds to kpc is based on the cosmology of H$_0$=70 km/s/Mpc, $\Omega_M$=0.27 and $\Omega_\Lambda$ = 0.73.}
\tablenotetext{3}{Source fell on chip gap on PN detector for ObsIDs 0303420101 and 0303420201; these spectra were not fit.}
\tablenotetext{4}{Dashes indicate that the spectra suffered from pile-up in that particular detector and were therefore not fit. See \citet{me3} for details.}
\end{deluxetable}
\end{landscape}

\clearpage
\begin{landscape}
\begin{deluxetable}{lcccccrr}

\tablewidth{0pt}
\tablecaption{\label{o3_apec}[OIII] Sample: APEC model parameters (solar abundance)}
\tablehead{
\colhead{Galaxy} & \colhead{N$_{H,1}$} & \colhead{kT$_1$} & \colhead{kT$_2$} & \colhead{$\Gamma$} &
\colhead{N$_{H,2}$}   & \colhead{$\chi^2$ 2 APECs} &  \colhead{$\chi^2$ 1 APEC} \\
\colhead{ }      & \colhead{10$^{22}$ cm$^{-2}$} & \colhead{keV} & \colhead{keV} & &
\colhead{10$^{22}$cm$^{-2}$} & \colhead{DOF} & \colhead{DOF}}

\startdata

Mrk 0609\tablenotemark{1} & 0.04 & 0.27$^{+0.05}_{-0.04}$ & ...& 1.77$^{+0.05}_{-0.04}$ & 
... &  ... & 160 (203) \\

IC 0486 & $<$0.06 & $<$0.16 & ... & 1.23$^{+0.08}_{-0.07}$ & 
1.00$^{+0.10}_{-0.09}$ & ... & 636 (629) \\

2MASX J08035923+2345201\tablenotemark{2} & 0.61$^{+0.10}_{-0.14}$ & $<$0.12 & ... & 2.84$^{+1.05}_{-1.18}$ & 
46.7$^{+22.6}_{-24.1}$ & ... & 120 (95) \\

2MASX J08244333+2959238\tablenotemark{1} & 0.03 & 0.68$^{+0.06}_{-0.06}$ & $<$0.12 & 1.54$^{+0.36}_{-0.34}$ & 
15.5$^{+2.8}_{-2.4}$ & 229 (186) & 263 (188)\\

2MASX J10181928+3722419\tablenotemark{1} & 0.01 & 0.18$^{+0.04}_{-0.05}$  & ... &  2.63$^{+0.64}_{-0.69}$ &
... & ... & 52.1 (61) \\

2MASX J11110693+0228477\tablenotemark{3}  & $<$0.62 & 0.23$^{+0.15}_{-0.09}$ & ... & 2.04$^{+1.88}_{-0.88}$ & 
... & ... & 37.5 (36) \\

CGCG 242-028\tablenotemark{2} & 0.69$^{+0.17}_{-0.23}$ & $<$0.15 & ... & 0.31$^{+0.46}_{-0.49}$ &
... & ... & 87.0 (90) \\

SBS 1133+572 & $<$0.10 & 0.824$^{+0.23}_{-0.21}$ & ... & 3.08$^{+0.61}_{-0.38}$ & 
57.6$^{+45.4}_{-30.2}$  & ... & 38.1 (48) \\

Mrk 1457 & 0.66$^{+0.12}_{-0.12}$ & 0.14$^{+0.03}_{-0.03}$  & ... & 1.29$^{+1.37}_{-1.14}$ &
27.5$^{+15.8}_{-9.3}$ & ... & 35.2 (35) \\

2MASX J11570483+5249036 & $<$0.1 & 0.17$^{+0.03}_{-0.05}$ & ... & 2.85$^{+0.23}_{-0.34}$ & 
83.9$^{+47.1}_{-27.9}$ & ... & 123 (76) \\

2MASX J12183945+4706275\tablenotemark{1} & 0.02 & $<$0.24  & ... & 1.95$^{+0.70}_{-0.86}$ & 
87.2$^{+66.8}_{-34.1}$  & ... & 15 (19) \\

2MASX J12384342+0927362 & $<$0.07 & $<$0.23 & ... & 2.17$^{+0.31}_{-0.22}$ & 
29.3$^{+3.1}_{-2.6}$ & ... & 195 (164) \\

NGC 5695 & $<$0.59 & 0.24$^{+0.73}_{-0.12}$ & ... & 2.55$^{+0.68}_{-0.52}$ & 
... &  ... & 75.7 (62) \\

\enddata

\tablenotetext{1}{Best fit absorption same as Galactic absorption. This parameter was then frozen to the Galactic value.}
\tablenotetext{2}{Used c-stat.}
\tablenotetext{3}{PN only, MOS1 and MOS2 had low signal-to-noise spectra.}
\end{deluxetable}

\end{landscape}

\clearpage
\begin{landscape}
\begin{deluxetable}{lcccccrr}
\footnotesize
\tablewidth{0pt}
\tablecaption{\label{12m_apec}APEC model parameters (solar abundance\tablenotemark{1})}
\tablehead{
\colhead{Galaxy} & \colhead{N$_{H,1}$} & \colhead{kT$_1$} & \colhead{kT$_2$} & \colhead{$\Gamma$} &
\colhead{N$_{H,2}$}   & \colhead{$\chi^2$ 2 APECs} &  \colhead{$\chi^2$ 1 APEC} \\
\colhead{ }      & \colhead{10$^{22}$ cm$^{-2}$} & \colhead{keV} & \colhead{keV} & &
\colhead{10$^{22}$cm$^{-2}$} & \colhead{DOF} & \colhead{DOF}}

\startdata

NGC 0424 & 0.04$^{+0.03}_{-0.03}$ & 0.69$^{+0.08}_{-0.05}$ & $<$0.13 & 2.21$^{+0.21}_{-0.28}$ & 
32.6$^{+8.5}_{-6.0}$ & 246 (178) & 267 (180)\\

NGC 1144\tablenotemark{1} & 0.06 & 0.37$^{+0.29}_{-0.06}$ & ... & 1.91$^{+0.37}_{-0.24}$ & 
47.0$^{+3.5}_{-3.2}$ & .. & 175 (149) \\

NGC 1320 & $<$0.08 & 0.15$^{+0.03}_{-0.03}$ & 0.73$^{+0.09}_{-0.02}$ & 2.97$^{+0.27}_{-0.21}$ &
48.0$^{+35.6}_{-15.5}$ & 233 (188) & 279 (190) \\

NGC 1386 & 0.04$^{+0.02}_{-0.01}$ & 0.67$^{+0.03}_{-0.03}$ & 0.13$^{+0.02}_{-0.01}$ & 2.72$^{+0.12}_{-0.14}$ & 
31.4$^{+22.9}_{-11.5}$ & 398 (332) & 435 (334) \\

NGC 1667\tablenotemark{2} & 0.05 & 0.33$^{+0.07}_{-0.04}$ & ... & 2.18$^{+0.34}_{-0.37}$ & 
... & ... & 49.8 (38) \\

F05189-2524\tablenotemark{2} & 0.02 & 0.1 & ... & 2.05$^{+0.14}_{-0.14}$ & 
6.77$^{+0.44}_{-0.42}$ & ... & 501 (374) \\

NGC 3982\tablenotemark{2} & 0.01 & 0.29$^{+0.03}_{-0.03}$ & ... & 2.39$^{+0.18}_{-0.15}$ &
40.3$^{+25.5}_{-16.3}$ & ... & 174 (160)  \\

NGC 4388 (XMM)\tablenotemark{2} & 0.03 & 0.60$^{+0.04}_{-0.03}$ & 0.15$^{+0.02}_{-0.03}$ & 1.25$^{+0.12}_{-0.12}$ &
24.3$^{+1.1}_{-1.0}$ & 510 (495) & 580 (498) \\
NGC 4388 (Chandra)\tablenotemark{2} & 0.03 & 0.60$^{+0.04}_{-0.04}$ & 0.15$^{+0.04}_{-0.02}$ & 0.38$^{+0.29}_{-0.30}$ &
25.6$^{+3.1}_{-2.9}$ & 210 (166) &  269 (168) \\

NGC 4501\tablenotemark{2} & 0.03 & 0.36$^{+0.05}_{-0.03}$ & ... & 1.52$^{+0.30}_{-0.29}$ & 
... & ... & 94.2 (102) \\

TOLOLO 1238-364\tablenotemark{3} & $<$0.19 & 0.32$^{+0.44}_{-0.06}$ & ... & 2.45$^{+0.33}_{-0.36}$ & ... & ... 
& 91.6 (105) \\

NGC 4968\tablenotemark{2,3} & 0.08 & 0.68$^{+0.12}_{-0.12}$ & ... & 1.72$^{+0.18}_{-0.21}$ & 
... & ... & 325 (268)\\

M-3-34-64\tablenotemark{2} & 0.05 & 2.85$^{+0.34}_{-0.38}$ & 0.77$^{+0.02}_{-0.02}$ & 3.24$^{+0.19}_{-0.23}$ &
53.6$^{+2.9}_{-3.6}$ & 780 (492) & 848 (493) \\

NGC 5135\tablenotemark{2} & 0.05 & 0.73$^{+0.03}_{-0.03}$ & ... & 2.34$^{+0.07}_{-0.08}$ & 
... & .. & 166 (114) \\

NGC 5194 & 0.04$^{+0.01}_{-0.01}$ & 0.18$^{+0.01}_{-0.01}$ & 0.62$^{+0.01}_{-0.01}$ & 3.18$^{+0.14}_{-0.14}$ &
10.2$^{+0.80}_{-0.75}$ & 1560 (1291) & 2021 (1293) \\

NGC 5347\tablenotemark{2,3} & 0.02 & $<$0.21 & ... & 1.53$^{+0.30}_{-0.29}$ & 
32.6$^{+24.1}_{-19.6}$ & ... & 69.9 (82)\\

Mrk 463\tablenotemark{2} & 0.02 & 0.74$^{+0.04}_{-0.02}$ & 0.20$^{+0.06}_{-0.04}$ & 1.76$^{+0.13}_{-0.16}$ & 
28.6$^{+4.5}_{-3.4}$ & 365 (316) & 392 (318) \\

NGC 5506\tablenotemark{4} & 0.11$^{+0.01}_{-0.01}$ & 0.74$^{+0.05}_{-0.05}$ & ... & 1.71$^{+0.02}_{-0.01}$ &
2.69$^{+0.02}_{-0.03}$ & ... & 2646 (2380) \\
NGC 5506\tablenotemark{5} & '' & 0.94$^{+0.39}_{-0.24}$ & '' & '' & 
'' & ''& ''  \\
NGC 5506\tablenotemark{6}& 0.17$^{+0.01}_{-0.01}$ & ... & ... & 1.83$^{+0.02}_{-0.02}$ & 
2.80$^{+0.01}_{-0.02}$ & ... & 3982 (3137) \\
NGC 5506\tablenotemark{7} & '' & '' & '' & 1.76$^{+0.01}_{-0.00}$ & 
'' & '' & '' \\
NGC 5506\tablenotemark{8} & 0.12$^{+0.01}_{-0.01}$ & 0.83$^{+0.03}_{-0.03}$ & '' & '' &
'' & '' & '' \\
NGC 5506\tablenotemark{9} & '' & 0.96$^{+0.05}_{-0.05}$ & '' & '' & 
'' & '' & '' \\

Arp 220\tablenotemark{1,2} & 0.04 & 0.79$^{+0.04}_{-0.04}$ & ... & 0.95$^{+0.26}_{-0.24}$ &
... & ... & 309(302) \\

NGC 6890\tablenotemark{3} & $<$0.22 & 0.78$^{+0.24}_{-0.19}$ & ... & 3.28$^{+0.88}_{-0.74}$ & 
27.4$^{+18.4}_{-11.3}$ & ... & 164 (148) \\

IC 5063\tablenotemark{2,10} & 0.06 & 0.60$^{+0.10}_{-0.11}$ & ... & 1.85$^{+0.16}_{-0.21}$ &
21.0$^{+1.1}_{-1.2}$ & ... & 198 (141) \\

NGC 7130\tablenotemark{3} & 0.06$^{+0.03}_{-0.02}$ & 0.63$^{+0.03}_{-0.03}$ & ... & 2.43$^{+0.25}_{-0.18}$ &
75.4$^{+55.6}_{-38.1}$ & ... & 334 (240) \\

NGC 7172\tablenotemark{2} & 0.02 & 0.28$^{+0.05}_{-0.04}$ & ... & 1.56$^{+0.02}_{-0.03}$ &
7.36$^{+0.10}_{-0.10}$ & ... & 2074 (1746) \\
NGC 7172\tablenotemark{2} & '' & 0.26$^{+0.02}_{-0.02}$ & ... & 1.54$^{+0.02}_{-0.03}$ & 
8.13$^{+0.11}_{-0.11}$ & '' & '' \\

NGC 7582 (Chandra)\tablenotemark{2} & 0.01 & 0.72$^{+0.05}_{-0.05}$ & ... & 1.94$^{+0.11}_{-0.10}$ & 
24.6$^{+1.8}_{-1.6}$ & ... & 355 (301) \\
NGC 7582 (XMM)\tablenotemark{2} & 0.01 & 0.18$^{+0.02}_{-0.04}$ & 0.72$^{+0.01}_{-0.01}$ & 1.75$^{+0.05}_{-0.03}$ &
27.0$^{+1.5}_{-1.5}$ & 1438 (886) & 1586 (888) \\

NGC 7674\tablenotemark{2,3} & 0.04 & $<$0.11 & 0.67$^{+0.07}_{-0.05}$ & 0.62$^{+0.47}_{-0.43}$ & 
89.0$^{+69.0}_{-40.2}$ & 66.6 (70) & 113 (72) \\

\enddata

\tablenotetext{1}{All abundances frozen to solar except for Arp 220 which has an abundance of 0.17$^{+0.12}_{-0.05}$.}
\tablenotetext{2}{Best fit absorption same as Galactic absorption. This parameter was then frozen to the Galactic value.}
\tablenotetext{3}{Used c-stat.}
\tablenotetext{4}{ObsIDs 0201830201, 0201830301 and 0201830401. APEC and first power law normalizations fit independently.}
\tablenotetext{5}{ObsID 0013140101.}
\tablenotetext{6}{ObsID 0201830501.}
\tablenotetext{7}{ObsID 0013140201.}
\tablenotetext{8}{ObsID 0554170201.}
\tablenotetext{9}{ObsID 0554170101.}
\tablenotetext{10}{Used pile-up model.}
\end{deluxetable}

\clearpage
\end{landscape}

\begin{deluxetable}{lccc}

\tablewidth{0pt}
\tablecaption{\label{o3_flux}[OIII] Sample Fluxes (10$^{-14}$ erg cm$^{-2}$ s$^{-1}$)}
\tablehead{

\colhead{Galaxy} & \colhead{F$_{0.5-2keV}$} & \colhead{F$_{APEC}$} & \colhead{F$_{pow}$} }

\startdata

Mrk 0609                 &  79.1$_{-3.62}^{+3.62}$ & 4.82$_{-2.15}^{+2.14}$  &  74.3$_{-2.92}^{+2.92}$   \\
IC 0486                  &  35.0$_{-8.84}^{+17.6}$ & 0.45$_{-0.39}^{+1.12}$  &  34.6$_{-8.83}^{+17.6}$   \\
2MASX J08035923+2345201  &  1.24$_{-0.71}^{+0.86}$ & 0.83$_{-0.70}^{+0.84}$  &  0.41$_{-0.14}^{+0.17}$   \\
2MASX J08244333+2959238  &  5.25$_{-1.58}^{+1.26}$ & 3.04$_{-1.47}^{+1.04}$  &  2.21$_{-0.59}^{+0.71}$   \\
2MASX J10181928+3722419  &  2.27$_{-0.62}^{+0.92}$ & 1.04$_{-0.43}^{+0.80}$  &  1.23$_{-0.44}^{+0.46}$   \\
2MASX J11110693+0228477  &  1.44$_{-0.59}^{+419}$  & 0.42$_{-0.36}^{+419}$   &  1.02$_{-0.46}^{+1.31}$   \\
CGCG 242-028             &  1.17$_{-0.75}^{+2.52}$ & 0.80$_{-0.73}^{+2.50}$  &  0.37$_{-0.19}^{+0.33}$   \\
SBS 1133+572             &  3.46$_{-0.82}^{+1.27}$ & 0.52$_{-0.35}^{+0.36}$  &  2.94$_{-0.74}^{+1.22}$   \\
Mrk 1457                 &  2.90$_{-1.82}^{+10.3}$ & 2.36$_{-1.79}^{+10.3}$  &  0.54$_{-0.32}^{+0.66}$   \\
2MASX J11570483+5249036  &  3.83$_{-0.66}^{+1.31}$ & 0.91$_{-0.41}^{+1.09}$  &  2.92$_{-0.51}^{+0.72}$   \\
2MASX J12183945+4706275  &  1.54$_{-0.54}^{+2.15}$ & 0.55$_{-0.31}^{+2.10}$  &  0.99$_{-0.44}^{+0.47}$   \\
2MASX J12384342+0927362	 &  6.23$_{-0.76}^{+2.81}$ & 0.86$_{-0.39}^{+2.58}$  &  5.37$_{-0.65}^{+1.12}$   \\
NGC 5695                 &  2.57$_{-0.90}^{+5.37}$ & 0.64$_{-0.56}^{+4.31}$  &  1.93$_{-0.71}^{+3.20}$   \\

\enddata
\end{deluxetable}

\begin{landscape}

\begin{deluxetable}{lccccl}

\tablewidth{0pt}
\tablecaption{\label{12m_flux}12$\mu$m Sample Fluxes (10$^{-14}$ erg cm$^{-2}$ s$^{-1}$)}
\tablehead{

\colhead{Galaxy} & \colhead{F$_{0.5-2keV}$} & \colhead{F$_{0.5-2keV,extended}$\tablenotemark{1}} &  \colhead{F$_{APEC}$} & \colhead{F$_{pow}$} & \colhead{Comments}}

\startdata

NGC 0424        & 31.5$_{-8.53}^{+12.5}$ &  ...                    & 10.6$_{-7.81}^{+12.1}$ &  20.9$_{-3.43}^{+3.26}$ \\  
NGC 1144        & 8.39$_{-1.61}^{+1.25}$ &  ...                    & 2.87$_{-1.39}^{+0.87}$ &  5.52$_{-0.81}^{+0.90}$ \\   
NGC 1320        & 28.4$_{-5.46}^{+15.8}$ &  ...                    & 10.5$_{-4.92}^{+15.6}$ &  17.9$_{-2.73}^{+2.38}$ \\  
NGC 1386        & 25.2$_{-3.99}^{+6.35}$ &  8.48$^{+1.41}_{-1.36}$ & 11.7$_{-3.83}^{+6.19}$ &  13.5$_{-1.12}^{+1.43}$ \\  
NGC 1667        & 7.31$_{-1.52}^{+1.45}$ &  ...                    & 3.22$_{-1.10}^{+1.03}$ &  4.09$_{-1.05}^{+1.01}$ \\    
F05189-2524     & 12.1$_{-1.00}^{+2.02}$ &  ...                    & 2.22$_{-0.68}^{+1.88}$ &  9.88$_{-0.74}^{+0.74}$ \\   
NGC 3982        & 9.14$_{-0.96}^{+0.90}$ &  7.47$^{+1.96}_{-2.49}$ & 3.39$_{-0.73}^{+0.63}$ &  5.75$_{-0.62}^{+0.64}$ \\    
NGC 4388        & 28.5$_{-7.0}^{+16.0}$  &  23.3$^{+6.6}_{-6.4}$   & 17.8$_{-6.6}^{+15.4}$  &  10.7$_{-2.18}^{+3.15}$ & {\it Chandra} \& {\it XMM-Newton} observations averaged\\    
NGC 4501        & 7.24$_{-1.30}^{+1.57}$ &  5.47$^{+1.42}_{-1.47}$ & 3.83$_{-0.95}^{+1.06}$ &  3.41$_{-0.88}^{+1.15}$ \\    
TOLOLO 1238-364 & 14.2$_{4.2}^{+49}$     &  ...                    & 5.43$_{-3.80}^{+5.01}$ &  8.77$_{-1.82}^{+49}$   \\
NGC 4968        & 5.85$_{-1.05}^{+1.14}$ &  ...                    & 1.27$_{-0.60}^{+0.67}$ &  4.58$_{-0.86}^{+0.93}$ \\     
M-3-34-64	& 50.8$_{-7.81}^{+6.37}$ &  ...                    & 33.4$_{-7.29}^{+4.72}$ &  17.4$_{-2.80}^{+4.28}$ \\   
NGC 5135        & 34.5$_{-1.68}^{+1.66}$ &  16.4$^{+7.1}_{-5.9}$   & 17.0$_{-1.27}^{+1.23}$ &  17.5$_{-1.10}^{+1.12}$ \\   
NGC 5194        & 69.6$_{-6.07}^{+6.82}$ &  56.1$^{+8.0}_{-7.9}$   & 52.7$_{-5.97}^{+6.71}$ &  16.9$_{-1.09}^{+1.19}$ \\   
NGC 5347	& 2.94$_{-0.46}^{+0.87}$ &  ...                    & 0.50$_{-0.30}^{+0.80}$ &  2.44$_{-0.35}^{+0.34}$ \\    
Mrk 463         & 14.3$_{-2.72}^{+2.68}$ &  6.04$^{+9.40}_{-1.88}$ & 7.50$_{-2.62}^{+2.59}$ &  6.80$_{-0.73}^{+0.70}$ \\    
NGC 5506        & 308$_{-53}^{+56}$      &  ...                    & 6.07$_{-4.55}^{+4.48}$ &  302$_{-53}^{+56}$      & ObsIDs 0013140101, 0201830201, 0201830301, 0201830401 \\
NGC 5506	& 444$_{-37}^{+37}$      &  ...                    & 0  	   	    &  444$_{-37}^{+37}$      & ObsIDs 0013140201 \& 0201830501 \\   
NGC 5506	& 457$_{-43}^{+37}$      &  ...                    & 10.5$_{-1.52}^{+1.65}$ &  446$_{-43}^{+37}$      & ObsIDs 0554170201 \& 0554170101 \\    
Arp 220         & 4.90$_{-1.34}^{+1.19}$ &  4.38$^{+1.11}_{-0.77}$ & 3.05$_{-1.24}^{+0.95}$ &  1.85$_{-0.51}^{+0.71}$ \\    
NGC 6890        & 11.5$_{-3.76}^{+6.19}$ &  ...                    & 2.62$_{-1.40}^{+1.64}$ &  8.88$_{-3.49}^{+5.97}$ \\   
IC 5063		& 22.9$_{-3.18}^{+3.12}$ &  4.79$^{+1.15}_{-1.20}$ & 4.57$_{-1.58}^{+1.58}$ &  18.3$_{-2.76}^{+2.69}$ \\   
NGC7130         & 23.6$_{-1.89}^{+2.63}$ &  8.55$^{+3.65}_{-1.20}$ & 10.3$_{-0.96}^{+1.24}$ &  13.3$_{-1.62}^{+2.32}$ \\   
NGC7172		& 20.3$_{-2.45}^{+2.28}$ &  ...                    & 2.05$_{-0.51}^{+0.50}$ &  18.3$_{-2.40}^{+2.22}$ & ObsID 0414580101 \\   
NGC7172		& 11.8$_{-0.72}^{+0.75}$ &  ...                    & 2.06$_{-0.28}^{+0.27}$ &  9.74$_{-0.67}^{+0.70}$ & ObsIDs 0147920601 \& 0202860101 \\   
NGC7582		& 40.3$_{-3.7}^{+3.2}$   &  30.8$^{+3.2}_{-2.8}$   & 15.2$_{-3.1}^{+2.5}$   &  25.1$_{-2.0}^{+2.0}$   & {\it Chandra} \& {\it XMM-Newton} observations averaged \\    
NGC7674		& 17.5$_{-4.95}^{+2.91}$ &  ...                    & 13.6$_{-4.74}^{+2.17}$ &  3.90$_{-1.42}^{+1.94}$ \\   

\enddata
\tablenotetext{1}{Sy2s observed with {\it Chandra} where the unresolved emission after removal of the AGN had high enough signal-to-noise for adequate spectral fitting. F$_{0.5-2keV,extended}$ flux corresponds to this extended emission from the host galaxy (i.e. with the AGN removed).}
\end{deluxetable}

\end{landscape}

\clearpage
\begin{landscape}
\begin{deluxetable}{lcccccrr}
\footnotesize
\tablewidth{0pt}
\tablecaption{\label{pt_src_rem}{\it Chandra} Spectral Fits to Unresolved Emission Only\tablenotemark{2}}
\tablehead{
\colhead{Galaxy} & \colhead{N$_{H,1}$} & \colhead{kT$_1$} & \colhead{kT$_2$} & \colhead{$\Gamma$} &
\colhead{N$_{H,2}$}   & \colhead{$\chi^2$ 2 APECs} &  \colhead{$\chi^2$ 1 APEC} \\
\colhead{ }      & \colhead{10$^{22}$ cm$^{-2}$} & \colhead{keV} & \colhead{keV} & &
\colhead{10$^{22}$cm$^{-2}$} & \colhead{DOF} & \colhead{DOF}}

\startdata

NGC 1386 & $<$0.05 & 0.58$^{+0.09}_{-0.13}$ & ... & 3.23$^{+0.38}_{-0.34}$ & 48.9$^{+32.0}_{-12.4}$ & ... & 79.0 (54) \\

NGC 3982\tablenotemark{2} & 0.01 & 0.20$^{+0.06}_{-0.06}$ & ... & 2.04$^{+0.68}_{-0.68}$ & ... & ... & 46.5 (25) \\

NGC 4388\tablenotemark{2} & 0.03 & 0.60$^{+0.04}_{-0.04}$ & 0.16$^{+0.04}_{-0.02}$ & 0.98$^{+0.51}_{-0.50}$ & ... & 85.4 (83) & 123 (84)\\

NGC 4501\tablenotemark{2} & 0.03 & 0.37$^{+0.22}_{-0.07}$ & ... & 2.26$^{+0.44}_{-0.50}$ & ... & ... & 31.2(38) \\

NGC 5135 & $<$0.13 & 0.84$^{+0.09}_{-0.09}$ & 0.39$^{+0.11}_{-0.09}$ & 2.24$^{+0.33}_{-0.31}$ & ... & 62.8 (60) & 77.9 (62) \\

NGC 5194\tablenotemark{2} & 0.02 & 0.60$^{+0.01}_{-0.01}$ & 0.18$^{+0.03}_{-0.02}$ & 3.13$^{+0.10}_{-0.09}$ & 7.92$^{+1.38}_{-1.08}$ &
525 (352) & 616 (353) \\

Mrk 463 & $<$0.05 & 0.62$^{+0.05}_{-0.07}$ & ... & 2.50$^{+0.75}_{-0.47}$ & 34.6$^{+10.5}_{-7.7}$ & ... & 86 (71) \\

Arp 220\tablenotemark{2} & $<$0.04 & 0.73$^{+0.07}_{-0.12}$ & ... & 2.10$^{+0.44}_{-0.36}$ & ... & ... & 69.3 (57) \\

IC 5063\tablenotemark{2} & 0.06 & 0.34$^{+0.09}_{-0.04}$ & ... & 1.51$^{+0.55}_{-0.59}$ & 28.5$^{+6.2}_{-5.0}$ & ... & 71.5 (57) \\

NGC 7130 & 0.05$^{+0.07}_{-0.03}$ & 0.47$^{+0.07}_{-0.10}$ & ... & 2.24$^{+0.37}_{-0.29}$ & ... & ... & 58.2 (50) \\

NGC 7582 & 0.03$^{+0.02}_{-0.02}$ & 0.73$^{+0.04}_{-0.05}$ & ... & 2.37$^{+0.23}_{-0.20}$ & 41.8$^{+10.1}_{-8.2}$ & ... & 142 (128) \\

\enddata

\tablenotetext{1}{All abundances frozen to solar.}
\tablenotetext{2}{Best fit absorption same as Galactic absorption. This parameter was then frozen to the Galactic value.}
\end{deluxetable}
\end{landscape}

% Begin figures here

\clearpage

\begin{deluxetable}{lccc}

\tablewidth{0pt}
\tablecaption{\label{o3_lum}[OIII] Sample Star-formation and AGN Luminosities}
\tablehead{

\colhead{Galaxy} & \colhead{Log(L$_{x,SF}$)} &  \colhead{Log(L$_{x,AGN}$)} }

\startdata

Mrk 0609                 & 41.38$^{+0.54}_{-0.39}$ & 42.28$^{+0.06}_{-0.06}$ \\
IC 0486                  & 40.39$^{+0.87}_{-0.42}$ & 41.83$^{+0.27}_{-0.24}$ \\
2MASX J08035923+2345201  & 40.45$^{+0.87}_{-0.33}$ & 39.64$^{+0.33}_{-0.30}$ \\
2MASX J08244333+2959238  & 40.81$^{+0.60}_{-0.33}$ & 40.18$^{+0.30}_{-0.30}$ \\
2MASX J10181928+3722419  & 40.99$^{+0.72}_{-0.36}$ & 40.54$^{+0.27}_{-0.24}$ \\
SBS 1133+572             & 40.76$^{+0.78}_{-0.33}$ & 41.17$^{+0.15}_{-0.15}$ \\
Mrk 1457                 & 41.74$^{+0.96}_{-0.42}$ & 40.15$^{+0.57}_{-0.48}$ \\
2MASX J11570483+5249036  & 40.67$^{+0.84}_{-0.36}$ & 40.81$^{+0.12}_{-0.12}$ \\
2MASX J12183945+4706275  & 41.68$^{+0.93}_{-0.42}$ & 41.20$^{+0.33}_{-0.30}$ \\
2MASX J12384342+0927362	 & 41.69$^{+0.87}_{-0.42}$ & 41.90$^{+0.15}_{-0.12}$ \\
NGC 5695                 & 40.78$^{+0.96}_{-0.81}$ & 39.79$^{+0.42}_{-0.41}$ \\

\enddata
\end{deluxetable}

\clearpage

\begin{landscape}
\begin{deluxetable}{lccl}

\tablewidth{0pt}
\tablecaption{\label{12m_lum}12$\mu$m Sample Star-formation and AGN Luminosities}
\tablehead{

\colhead{Galaxy} & \colhead{Log(L$_{x,SF}$)} &  \colhead{Log(L$_{x,AGN}$)} & \colhead{Comments}}

\startdata

NGC 0424        & 40.99$^{+0.69}_{-0.54}$ & 40.72$^{+0.21}_{-0.18}$ \\
NGC 1144        & 40.78$^{+0.66}_{-0.24}$ & 40.87$^{+0.21}_{-0.21}$ \\
NGC 1320        & 40.33$^{+1.05}_{-0.51}$ & 40.42$^{+0.21}_{-0.21}$ \\
NGC 1386\tablenotemark{1} & 39.23$^{+0.07}_{-0.08}$ & 39.52$^{+0.10}_{-0.07}$ \\
NGC 1667        & 40.30$^{+0.60}_{-0.21}$ & 40.06$^{+0.24}_{-0.24}$ \\
F05189-2524     & 41.20$^{+0.78}_{-0.30}$ & 41.56$^{+0.15}_{-0.12}$ \\
NGC 3982\tablenotemark{1} & 39.42$^{+0.10}_{-0.18}$ & 38.77$^{+0.04}_{-0.05}$ \\
NGC 4388\tablenotemark{1} & 40.52$^{+0.11}_{-0.14}$ & 39.87$^{+0.19}_{-0.12}$  \\
NGC 4501\tablenotemark{1} & 39.89$^{+0.10}_{-0.14}$ & 39.40$^{+0.09}_{-0.09}$  \\
TOLOLO 1238-364 & 40.63$^{+0.72}_{-0.57}$ & 40.31$^{+0.48}_{-0.45}$ \\
NGC 4968        & 39.67$^{+0.69}_{-0.33}$ & 39.88$^{+0.12}_{-0.12}$ \\
M-3-34-64	& 41.47$^{+0.45}_{-0.21}$ & 40.84$^{+0.60}_{-0.60}$ \\
NGC 5135\tablenotemark{1} & 40.86$^{+0.16}_{-0.19}$ & 40.90$^{+0.02}_{-0.02}$ \\
NGC 5194\tablenotemark{1} & 39.69$^{+0.06}_{-0.07}$ & 39.08$^{+0.04}_{-0.04}$ \\
NGC 5347	& 39.34$^{+0.84}_{-0.48}$ & 39.47$^{+0.15}_{-0.12}$ \\
Mrk 463\tablenotemark{1} & 41.55$^{+0.41}_{-0.16}$ & 41.69$^{+0.07}_{-0.09}$ \\
NGC 5506        & 39.88$^{+0.93}_{-0.27}$ & 41.38$^{+0.12}_{-0.09}$  & ObsIDs 0013140101, 0201830201, 0201830301, 0201830401 \\
NGC 5506	& 0                       & 41.53$^{+0.09}_{-0.06}$  & ObsIDs 0013140201 \& 0201830501 \\   
NGC 5506	& 40.06$^{+0.45}_{-0.21}$ & 41.53$^{+0.06}_{-0.06}$  & ObsIDs 0554170201 \& 0554170101 \\    
Arp 220\tablenotemark{1} & 40.51$^{+0.10}_{-0.08}$ & 39.58$^{+0.09}_{-0.14}$  \\
NGC 6890        & 39.83$^{+0.69}_{-0.39}$ & 39.94$^{+0.24}_{-0.24}$ \\
IC 5063\tablenotemark{1} & 40.11$^{+0.09}_{-0.13}$ & 40.69$^{+0.06}_{-0.06}$ \\
NGC 7130\tablenotemark{1} & 40.69$^{+0.15}_{-0.07}$ & 40.94$^{+0.05}_{-0.04}$ \\
NGC 7172	& 39.85$^{+0.39}_{-0.36}$ & 40.45$^{+0.06}_{-0.06}$ & ObsID 0414580101 \\   
NGC 7172	& 39.67$^{+0.45}_{-0.15}$ & 40.18$^{+0.12}_{-0.12}$ & ObsIDs 0147920601 \& 0202860101 \\   
NGC 7582\tablenotemark{1} & 40.23$^{+0.04}_{-0.04}$ & 39.72$^{+0.03}_{-0.04}$ \\
NGC 7674	& 41.53$^{+0.51}_{-0.21}$ & 40.03$^{+2.70}_{-0.39}$ \\

\enddata
\tablenotetext{1}{Sy2s observed with {\it Chandra} where the AGN was removed and extended emission was fitted. L$_{x,SF}$ is derived from these flux values for these objects while L$_{x,AGN}$ is simply L$_{0.5-2keV}$-L$_{x,SF}$. For the remaining sources, L$_{x,SF}$ and L$_{x,AGN}$ is derived as described in the text.}
\end{deluxetable}

\end{landscape}
\clearpage

\end{document}